\RequirePackage{lineno}
\documentclass[aps,prl,twocolumn,groupedaddress,amsmath,amssymb]{revtex4-1}

\usepackage{ulem}
\usepackage{xcolor}
\usepackage{amssymb}
\usepackage{graphicx}
\usepackage{dcolumn}
\usepackage{bm}
\usepackage{color}
\usepackage{subfigure}

\begin{document}
\linenumbers
\newcommand{\jpsi}{J/\psi}
\newcommand{\pip}{\pi^+}
\newcommand{\pin}{\pi^-}
\newcommand{\pio}{\pi^0}
\newcommand{\g}{\gamma}
\newcommand{\gev}{GeV/c$^2$}
\newcommand{\mev}{MeV/c$^2$}
\newcommand{\ar}{\rightarrow}
\hyphenation{
Chromo-dy-nam-ics states studies im-port-ant using since decay in-ter-pret-ations
pres-ence
cal-or-im-eter inter-leaved reso-lution its
MDC EMC
hy-poth-esis
evi-dent
Lund-charm data add-ition-al ef-fi-ciencies
back-grounds de-tect-or simu-la-tion
stat-is-tic-al
Ref sigma error limit table upper
com-pari-sons
com-pari-son
study
among
BESIII
decays
con-sider
res-on-ance
BESII
meas-ure-ment
uni-ver-sity
}

\title{\Large \boldmath \bf Observation of a structure at 1.84 GeV/c$^2$ in the $3(\pi^+\pi^-)$ mass spectrum in $\jpsi\rightarrow \gamma 3(\pip\pin)$ decays}

\author{
\begin{small}
M.~Ablikim$^{1}$, M.~N.~Achasov$^{6,a}$, O.~Albayrak$^{3}$, D.~J.~Ambrose$^{39}$, F.~F.~An$^{1}$, Q.~An$^{40}$, J.~Z.~Bai$^{1}$, R.~Baldini Ferroli$^{17A}$, Y.~Ban$^{26}$, J.~Becker$^{2}$
% ...
\\
\vspace{0.2cm}
(BESIII Collaboration)\\
\vspace{0.2cm} {\it
$^{1}$ Institute of High Energy Physics, Beijing 100049, People's Republic of China\\
$^{2}$ Bochum Ruhr-University, D-44780 Bochum, Germany\\
$^{3}$ Carnegie Mellon University, Pittsburgh, Pennsylvania 15213, USA\\
% ...
$^{6}$ G.I. Budker Institute of Nuclear Physics SB RAS (BINP), Novosibirsk 630090, Russia\\
% ...
$^{17}$ (A)INFN Laboratori Nazionali di Frascati, I-00044, Frascati, Italy; (B)INFN and University of Perugia, I-06100, Perugia, Italy\\
% ...
$^{26}$ Peking University, Beijing 100871, People's Republic of China\\
% ...
$^{39}$ University of Rochester, Rochester, New York 14627, USA\\
$^{40}$ University of Science and Technology of China, Hefei 230026, People's Republic of China\\
% ...
$^{a}$ Also at the Novosibirsk State University, Novosibirsk, 630090, Russia\\
% ...
}
\end{small}
}

\vspace{0.4cm}

\begin{abstract}
With a sample of 225.3 million $J/\psi$ events taken with the BESIII
detector, the decay $\jpsi\rightarrow \gamma 3(\pip\pin)$ is
analyzed. A structure at 1.84 GeV/c$^2$ is observed in the $3(\pip\pin)$
invariant mass spectrum with a statistical
significance of 7.6$\sigma$ . The mass and width are measured to be
$M=1842.2\pm 4.2^{+7.1}_{-2.6}$ MeV/c$^2$ and $\Gamma=83\pm 14 \pm
11$ MeV. The product branching fraction is
determined to be $B(\jpsi\ar\gamma X(1840))\times B(X(1840)\ar
3(\pip\pin))=(2.44\pm0.36^{+0.60}_{-0.74})\times 10^{-5}$. No
$\eta^\prime$ signals are observed in the $3(\pip\pin)$ invariant
mass spectrum, and the upper limit of the branching fraction for the decay
$\eta^\prime\ar 3(\pip\pin)$ is set to be $3.1\times10^{-5}$ at a 90\% confidence level.
\end{abstract}


\maketitle


 Within the framework of Quantum Chromodynamics (QCD), the existence
of gluon self-coupling suggests that in addition to conventional meson
and baryon states, there may exist bound states such as glueballs,
hybrid states and multiquark states. Experimental searches for
glueballs and hybrid states have been carried out for many years, and
so far no conclusive evidence has been found.  The establishment of
new forms of hadronic matter beyond simple quark-antiquark system
remains one of the main interests in experimental particle physics.

Decays of the $J/\psi$ particle have always been regarded as an ideal
environment in which to study light hadron spectroscopy and search for
new hadrons.  At BESII, important advances in light hadron
spectroscopy were made using studies of $J/\psi$ radiative decays
~\cite{Bai:2003sw,Ablikim:2005um,Ablikim:2006dw}. Of interest is the
observation of the $X(1835)$ state in
$J/\psi\rightarrow\gamma\pi^+\pi^-\eta^\prime$ decay, which was
confirmed recently by BESIII~\cite{Ablikim:2010au} 
and CLEO-c~\cite{Alexander:2010vd}.  Since the discovery of the $X(1835)$, many possible
interpretations have been proposed, such as a $p\bar{p}$ bound
state~\cite{Ding:2005ew,Dedonder:2009bk,Liu:2007tj,Wang:2006sna}, a
glueball~\cite{Li:2005vd,Kochelev:2005vd}, or a radial excitation of
the $\eta^\prime$ meson~\cite{Huang:2005bc,Yu:2011ta}.  In the search
for the $X(1835)$ in other $J/\psi$ hadronic decays, BESIII reported
the first observation of the $X(1870)$ in
$J/\psi\rightarrow\omega\pi^+\pi^-\eta$~\cite{Ablikim:2011pu}.  More
recently, BESIII performed spin-parity analyses of threshold
structures, the $X(p\bar{p})$, observed in $J/\psi \to \gamma p
\bar{p}$~\cite{BESIII:2011aa}, and the
$X(1810)$, observed in
$J/\psi\rightarrow\gamma\omega\phi$~\cite{Ablikim:2012ft}.  The spin-parity of the
$X(p\bar{p})$ is found to be $0^{-+}$ and the $X(1810)$ is confirmed
to be a $0^{++}$ state.  To understand their nature, further study is
strongly needed, in particular, in searching for new decay modes.

Since the $X(1835)$ was confirmed to be a pseudoscalar particle~\cite{Ablikim:2010au} and it may have properties in common with
the $\eta_c$. Six charged pions is a known decay mode of the $\eta_c$; therefore,
$J/\psi$ radiative decays to $3(\pi^+\pi^-)$ may be a favorable channel to search for the
$X$ states in the 1.8 - 1.9 GeV/c$^2$ region.

In this letter, we present results of a study of $\jpsi\ar\gamma3(\pip\pin)$ decays
using a sample of $(225.3\pm2.8)\times 10^6$ $\jpsi$
events~\cite{Ablikim:2012cn} collected with the BESIII
detector~\cite{Ablikim:2009aa}. A structure at 1.84 GeV/c$^2$
(denoted as $X(1840)$ in this letter), is clearly observed in the mass spectrum of six
charged pions. Meanwhile in an attempt to search for
$\eta^\prime$ decaying into six charged pions, no $\eta^\prime$ signals are observed. The
upper limit on the decay branching fraction is set at a 90\% confidence level.

The BESIII detector is a magnetic spectrometer located at BEPCII~\cite{Bai:2001dw},
 a double-ring $e^+e^-$ collider with the design peak luminosity of $10^{33}
~\rm{cm}^{-2}\rm{s}^{-1}$ at a center of mass energy of 3.773 GeV.
The cylindrical core of the BESIII detector consists of a helium-based main drift
chamber (MDC), a plastic scintillator time-of-flight system (TOF), and
a CsI(Tl) electromagnetic calorimeter (EMC), which are all enclosed in
a superconducting solenoidal magnet providing a 1.0~T magnetic
field. The solenoid is supported by an octagonal flux-return yoke with
resistive plate counter muon identifier modules interleaved with
steel. The acceptance of charged particles and photons is 93\% over
4$\pi$ solid angle, and the charged-particle momentum
resolution at 1~GeV/c is 0.5\%.
The EMC measures photon energies with the resolution of
2.5\% (5\%) at 1 GeV in the barrel (endcaps).

Monte Carlo (MC) simulations are used to estimate the backgrounds and determine the detection efficiency.
Simulated events are processed using \textsc{geant}{\footnotesize4}~\cite{Agostinelli:2002hh,geant42},
where measured detector resolutions are incorporated.

Charged tracks are reconstructed using hits in the MDC
and are required to pass within $\pm$10 cm from the interaction point in the
beam direction and $\pm$1 cm in the perpendicular plane to the beam.
The polar angle of the charged
tracks should be in the region $|\cos\theta|<0.93$. Photon
candidates are selected from showers in the EMC with the energy deposit
in the EMC barrel region ($|\cos\theta|<0.8$) greater than 25 MeV and
in the EMC endcap region ($0.86<|\cos\theta|<0.92$) greater than
50 MeV. The photon candidates should be isolated
from the charged tracks by an opening angle of 10$^{\circ}$.

Candidate events are required to have six charged tracks
with zero net charge and at least one photon.
All the charged tracks are assumed to be pions.
The candidate events are required to successfully pass a primary
vertex fit. A four-momentum constraint (4C) kinematic fit is
performed to the $\jpsi\ar\gamma3(\pip\pin)$ hypothesis, and the
$\chi^2_{4C}$ is required to be less than 30. If the number of photon
candidates is more than one, the $\gamma3(\pip\pin)$ combination with the minimum
$\chi^2_{4C}$ is selected. To suppress background
events with multi-photons in the final states, $P_{t\gamma}^2 =
2|\vec{P}_{\text{miss}}|^2(1-\cos\theta_{\text{miss}})$
is required to
be less than 0.0004 GeV$^2$/c$^2$, where $\vec{P}_{\text{miss}}$ is the
missing momentum of the six charged tracks and $\theta_{\text{miss}}$
is the angle between the missing momentum and the momentum of the radiative photon.
To further reject backgrounds with additional photons in the final state,
the $\chi^2_{4C}$ of four-constraint kinematic fit in the hypothesis of $J/\psi\rightarrow\gamma 3(\pi^+\pi^-)$
is required to be less than that of the $\gamma\gamma 3(\pi^+\pi^-)$ hypothesis,
and the $\gamma\gamma$ invariant mass in the $\gamma\gamma3(\pi^+\pi^-)$ hypothesis
is required to be $|M(\gamma\gamma) - M(\pi^0)|>0.01$ GeV/c$^2$.
To suppress background events with $K_S\rightarrow\pi^+\pi^-$ in the final state,
$K_{S}$ candidates are reconstructed from secondary vertex fits to
all oppositely charged track pairs. The invariant mass $M(\pi^+\pi^-)$
must be within the range $|M(\pi^+\pi^-)-M(K_{S})|<0.005$ GeV/$c^2$,
where the $M(K_{S})$ is the nominal $K_{S}$ mass ~\cite{Beringer:1900zz}.
The number of $K_S$ candidates is required to be less than 2.

\begin{figure}
    \includegraphics[width=0.5\textwidth]{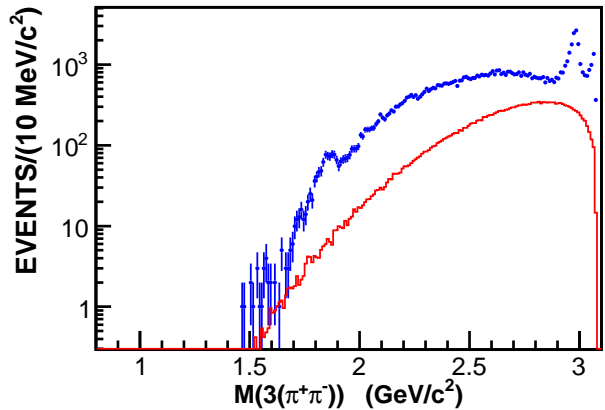}
    \caption{Distribution of the invariant mass of $3(\pip\pin)$ from $J/\psi\rightarrow\gamma 3(\pi^+\pi^-)$ events. The dots with error bars are data;
    the histogram is phase space events with an arbitrary normalization.
    \label{m6pi}}
\end{figure}

\begin{figure}
    \includegraphics[width=0.5\textwidth]{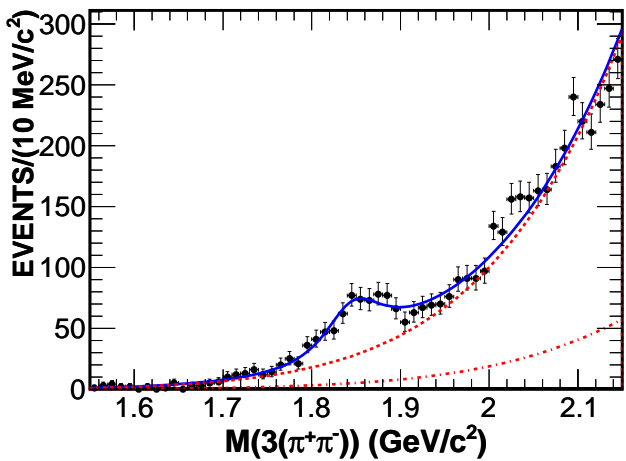}
    \caption{The fit of mass spectrum of $3(\pip\pin)$. The dots with error bars are data; the solid line is the fit result. The dashed line represents all the backgrounds, including the background events from $\jpsi\ar\pio3(\pip\pin)$ (dash-dotted line, fixed in the fit) and a third-order polynomial representing other backgrounds.\label{m6pifit}}
\end{figure}

Figure~\ref{m6pi} shows the $3(\pi^+\pi^-)$ invariant mass spectrum
for events that survive the above selection criteria, where a clear
$\eta_c$ peak is observed around 2.98 GeV/c$^2$, no evident
$\eta^\prime$ signal is observed, and a distinct enhancement is seen
around 1.84 GeV/c$^2$.  In Fig.~\ref{m6pifit}, the $M(3(\pi^+\pi^-))$
distribution is plotted in the range [1.55, 2.15] GeV/$c^2$.

To investigate possible backgrounds, we use a MC sample of 225
million simulated $J/\psi$ decays, in which the decays with known
branching fractions~\cite{Beringer:1900zz} are generated by
{\sc{BesEvtGen}}~\cite{Ping:2008zz} and unmeasured $J/\psi$ decays by
the Lundcharm model~\cite{Chen:2000tv}.  With the same selection
criteria, we find no evident structure at 1.84 GeV/$c^2$.
The background resulting from other, incorrectly reconstructed event
topologies is mainly from $J/\psi\rightarrow\pi^03(\pi^+\pi^-)$, which show no structure at 1.84 GeV/c$^2$
in the $3(\pi^+\pi^-)$ mass spectrum.
To estimate this contribution, we reconstruct the $\jpsi\ar\pi^03(\pip\pin)$
decay from data and then re-weight the $3(\pi^+\pi^-)$ invariant mass
spectrum by a multiplicative weighting factor
$\varepsilon_1/\varepsilon_2$, where $\varepsilon_1$ and
$\varepsilon_2$ are the efficiencies for
$J/\psi\rightarrow\pi^03(\pi^+\pi^-)$ MC events to pass
$J/\psi\rightarrow\gamma3(\pi^+\pi^-)$ and $J/\psi\rightarrow
\pi^03(\pi^+\pi^-)$ selection criteria, respectively.
The selection criteria for $\jpsi\ar\pi^03(\pip\pin)$ are similar to
those applied to $\jpsi\ar\gamma3(\pip\pin)$ except for the
requirement of an additional photon.  The background analysis shows
that the structure at 1.84 GeV/c$^2$ in the $3(\pi^+\pi^-)$ mass
spectrum does not come from background events.

To extract the number of signal events associated with the peaking structure, an unbinned maximum likelihood
fit is applied to the six pion mass spectrum.
The fit includes three components: a signal shape, shapes for the
$J/\psi\rightarrow\pi^03(\pi^+\pi^-)$ background and other
backgrounds, which have the same final states, but not contribute to
the structure around 1.84 GeV/c$^2$.  The signal shape is described
with a Breit-Wigner function modified by the effects of the phase
space factor and the detection efficiency, which is determined by a
phase-space MC simulation of $J/\psi\rightarrow\gamma 3(\pi^+\pi^-)$.
The Breit-Wigner function is convolved with a Gaussian function to
account for the detector resolution (5.1 MeV/c$^2$, determined from MC
simulation).
For the background shape, the contribution from the
$J/\psi\rightarrow\pi^03(\pi^+\pi^-)$ background, which is fixed in
the fit and shown by the dash-dotted line in Fig.~\ref{m6pifit}, is
represented by the re-weighted $3(\pi^+\pi^-)$ invariant mass
spectrum, while other contributions are represented by a third-order
polynomial. The total background is shown as the dashed line in
Fig.~\ref{m6pifit}.

The fit yields 632$\pm$93 events in the peak at 1842.2$\pm$4.2
MeV/c$^2$ and a width of $\Gamma$=83$\pm$14 MeV. The statistical
significance of the signal is determined from the change in log
likelihood and the change of number of degrees of freedom (d.o.f) in
the fit with and without the structure $X(1840)$.  Different
possibilities have been studied by varying the fit range and the
background shapes and by removing the phase space factor. Among all
possibilities the smallest statistical significance was 7.6$\sigma$
corresponding to $-2\Delta$lnL=67 and $\Delta$d.o.f=3. With the
detection efficiency, (11.5$\pm$0.1)\%, obtained from the phase space
MC simulation, the product branching fraction is measured to be
$B(\jpsi\ar\gamma X(1840))\times B(X(1840)\ar3(\pip\pin)) = (2.44 \pm
0.36)\times 10^{-5}$, where the error is statistical only.

No $\eta^\prime$ events are observed in the $3(\pip\pin)$ mass
spectrum. The upper limit at the 90\% confidence level is 2.44 events
with the confidence intervals suggested in
Ref.~\cite{Feldman:1997qc}. The detection efficiency 
in the mass region [0.928, 0.988] GeV/c$^2$ is determined to be
$(7.8\pm0.1)$\% from the MC simulation.  Since only the statistical
error is considered when we obtain the 90\% upper limit of the number
of events, the upper limit of the number of events
is shifted up by one sigma of the total
systematic uncertainty 
shown below in Table~\ref{system}. With the number of $J/\psi$ events and the
measured
$B(J/\psi\rightarrow\gamma\eta^\prime)=(5.16\pm0.15)\times10^{-3}$~\cite{Beringer:1900zz},
the upper limit of the branching fraction is obtained to be
$B(\eta^\prime\ar 3(\pip\pin))<3.1\times10^{-5}$.

Sources of systematic errors and their corresponding
contributions to the measurement of the branching fractions are
summarized in Table~\ref{system}. The uncertainties in
tracking and photon detection have been studied~\cite{Ablikim:2011kv} and the
difference between data and MC is about 2\% per charged track and 1\%
per photon, which is taken as the systematic error.
Uncertainty associated with the 4C kinematic fit comes from the inconsistency between data and MC simulation of the fit; this
difference is reduced by correcting the track helix parameters of MC simulation, as described in detail in Ref.~\cite{Ablikim:2012pg}.
In this analysis, we take the efficiency with correction as the nominal value, and take the difference between the
efficiencies with and without correction as the systematic uncertainty from the kinematic fit.
The background uncertainty is
determined by changing the background functions and the fit range.
The uncertainties from the mass spectrum fit include contributions from the
variation of the phase space factor and the possible impact
of other resonances (eg. $f_2(2010)$).~
The systematic error for the $P^2_{t\gamma}$ selection criterion is estimated
with the sample of $J/\psi\rightarrow\pi^03(\pi^+\pi^-)$ by comparing the efficiency
of this requirement between MC and data.
For the detection efficiency uncertainty due to the unknown spin-parity of the structure, we use the difference between phase space and a
pseudoscalar meson hypothesis.
The uncertainties from MC statistics,
the branching fraction of $J/\psi\rightarrow\gamma\eta^\prime$~\cite{Beringer:1900zz}
and the flux of $J/\psi$ events~\cite{Ablikim:2012cn} are also
considered. We assume all of these sources are independent,
and take the total systematic error to be their sum in quadrature.

The systematic uncertainties on mass and width are estimated from
the mass scale, background shape, fitting range, mass spectrum fit,
and possible biases due to the fitting procedure.
The uncertainty from the detector resolution is checked by using a double Gaussian function as the resolution function, and the change is found to be negligible. The uncertainty from the mass scale is estimated by fitting the $\eta_c$ resonance in $M(3(\pi^+\pi^-))$ spectrum.
Uncertainties from the background shape and fitting range are estimated by
varying the functional form used to represent the background and the fitting range.
Uncertainties from mass spectrum fit include contributions from
the variation of the phase space factor and the possible impact of
other resonances (eg. $f_2(2010)$).
Possible biases due to the fitting procedure are estimated from differences between the input and output
of the mass and width values from MC studies. Adding these sources in
quadrature, the total systematic error on the mass is
$^{+7.1}_{-2.6}$ MeV/c$^2$ and on the width is $\pm 11$ MeV.

\begin{table}[htpb]
\begin{center}
\caption{Summary of the systematic uncertainties in the branching fractions (in unit of \%).}\label{system}
\begin{tabular}{c|c|c}
\hline
Sources       &     $X(1840)$ & $\eta^\prime$ \\
\hline
MDC tracking & 12 &12 \\
\hline
Photon detection & 1 & 1 \\
\hline
$P^2_{t\gamma}$ cut & 2.0 & 2.0 \\
\hline
Kinematic fit &  4.3 & 5.1\\
\hline
Background uncertainty& 17.1 & - \\
\hline
Mass spectrum fit &$^{+10.3}_{-20.3}$ & -  \\
\hline
Detection efficiency &  6.1& -\\
\hline
MC statistics & 0.9 &  1.3 \\
\hline
$B(\jpsi\ar\g\eta^\prime)$ & -& 2.9\\
\hline
Number of $\jpsi$ events & 1.2 & 1.2\\
\hline
Total & $^{+24.6}_{-30.2}$ & 13.7 \\
\hline
\end{tabular}
\end{center}
\end{table}

\begin{figure}
    \centering
    \includegraphics[width=0.50\textwidth]{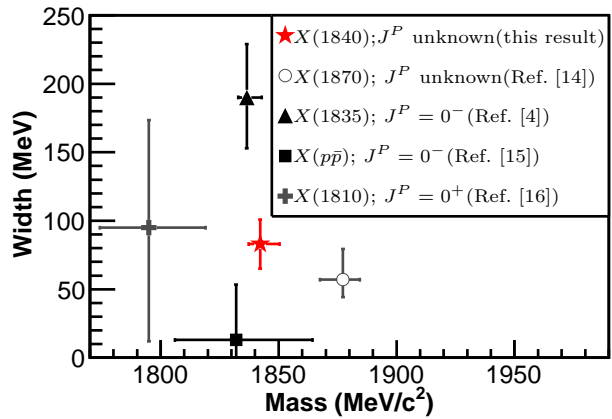}
    \put (-145,145) {\scriptsize{$X(1840)$;$J^P$ unknown(this result)}}
    \put (-145,130) {\scriptsize{$X(1870)$; $J^P$ unknown(Ref.~\cite{Ablikim:2011pu})}}
    \put (-145,115) {\scriptsize{$X(1835)$; $J^P=0^-$(Ref.~\cite{Ablikim:2010au})}}
    \put (-145,100) {\scriptsize{$X(p\bar{p})$; $J^P=0^-$(Ref.~\cite{BESIII:2011aa})}}
    \put (-145,85) {\scriptsize{$X(1810)$; $J^P=0^+$(Ref.~\cite{Ablikim:2012ft})}}
    \caption{Comparisons of observations at BESIII. The error bars
    include statistical, systematic, and, where applicable, model
    uncertainties.}  \label{compmw}
\end{figure}

In summary, we studied the decay $\jpsi\ar\g3(\pip\pin)$ with a 225.3
million $J/\psi$ event sample~\cite{Ablikim:2012cn} accumulated at the
BESIII detector.  A structure at 1.84 GeV/c$^2$ is observed in the
$3(\pip\pin)$ mass spectrum with a statistical significance of
7.6$\sigma$.  Fitting the structure $X(1840)$ with a modified
Breit-Wigner function yields $M=1842.2\pm4.2^{+7.1}_{-2.6}$ MeV/c$^2$
and $\Gamma=83\pm14\pm 11 $ MeV.  The product branching fraction is
determined to be $B(\jpsi\ar\g X(1840))\times
B(X(1840)\ar3(\pip\pin))=(2.44\pm 0.36^{+0.60}_{-0.74})\times
10^{-5}.$
The comparison to the BESIII results of the masses
and widths of the $X(1835)$~\cite{Ablikim:2010au},
$X(p\bar{p})$~\cite{BESIII:2011aa}, $X(1870)$~\cite{Ablikim:2011pu},
and $X(1810)$~\cite{Ablikim:2012ft} are displayed in
Fig.~\ref{compmw}, where the mass of $X(1840)$ is in
agreement with those of $X(1835)$ and $X(p\bar{p})$, while its width
is significantly different from either of them.
However, we do not include the BESII result
in Fig. 3 as a more precise study of the $X(1835)$ in BESIII~\cite{Ablikim:2010au} indicates that one must consider the presence of additional resonances above 2 GeV/c$^2$ that were not apparent in the BESII analysis to obtain an accurate determination of the width of the $X(1835)$.
Therefore, based on these data, one cannot determine whether $X(1840)$ is a new state or the signal of a $3(\pi^+\pi^-)$ decay mode of an existing state.  Further study, including an amplitude analysis to determine the spin and parity of the $X(1840)$, is needed to establish the relationship between different experimental observations in this mass region and determine the nature of the underlying resonance or resonances.

A search for $\eta^\prime\rightarrow 3(\pi^+\pi^-)$ is also
performed, but no $\eta^\prime$ signal is observed.
The upper limit on the branching fraction for the decay at
the 90\% confidence level is $B(\eta^\prime\rightarrow
3(\pi^+\pi^-))<3.1\times10^{-5}$, which is improved by one order of
magnitude compared to the previous measurement~\cite{cleo:2009}.

The BESIII collaboration thanks the staff of BEPCII and the computing center for their 
strong support.
This work is supported in part by the Ministry of Science and Technology of China under Contract No. 2009CB825200; National Natural Science Foundation of China (NSFC) under Contracts Nos. 10625524, 10821063, 10825524, 10835001, 10935007, 11125525, 11175189, 11235011; Joint Funds of the National Natural Science Foundation of China under Contracts Nos. 11079008, 11179007; the Chinese Academy of Sciences (CAS) Large-Scale Scientific Facility Program; CAS under Contracts Nos. KJCX2-YW-N29, KJCX2-YW-N45; 100 Talents Program of CAS; German Research Foundation DFG under Contract No. Collaborative Research Center CRC-1044; Istituto Nazionale di Fisica Nucleare, Italy; Ministry of Development of Turkey under Contract No. DPT2006K-120470; U. S. Department of Energy under Contracts Nos. DE-FG02-04ER41291, DE-FG02-05ER41374, DE-FG02-94ER40823; U.S. National Science Foundation; University of Groningen (RuG) and the Helmholtzzentrum fuer Schwerionenforschung GmbH (GSI), Darmstadt; 
National Research Foundation of Korea Grant No. 2011-0029457
and WU Grant No. R32-10155.



\end{document}